\documentclass[10pt,twocolumn,a4paper,twoside]{article}

\usepackage[a4paper,
left=0.65in,
right=0.65in,
top=0.8in,
bottom=0.7in]{geometry}

\usepackage{times}
\usepackage{amsmath,amssymb}
\usepackage{graphicx}
\usepackage{float}
\usepackage{subcaption}
\usepackage{caption}
\usepackage{booktabs}
\usepackage{multirow}
\usepackage{hyperref}
\usepackage[nameinlink,noabbrev]{cleveref}
\usepackage{url}
\usepackage[numbers]{natbib}
\usepackage{setspace}
\usepackage{xcolor}
\usepackage{orcidlink}
\usepackage{titlesec}

\titleformat{\section}
{\fontsize{12}{15}\selectfont\bfseries}
{\thesection}
{0.8em}
{\MakeUppercase}

\hypersetup{
colorlinks=true,
linkcolor=blue,
citecolor=blue,
urlcolor=blue
}

\crefname{figure}{Fig.}{Figs.}
\Crefname{figure}{Figure}{Figures}

\crefname{table}{Table}{Tables}
\Crefname{table}{Table}{Tables}

\crefname{equation}{Eq.}{Eqs.}
\Crefname{equation}{Equation}{Equations}

\crefname{section}{Section}{Sections}
\Crefname{section}{Section}{Sections}

\usepackage{fancyhdr}

\definecolor{orcidgreen}{HTML}{A6CE39}

\let\oldorcidlink\orcidlink
\renewcommand{\orcidlink}[1]{%
\textcolor{orcidgreen}{\oldorcidlink{#1}}
}

\renewcommand{\thefootnote}{\fnsymbol{footnote}}

\begin{document}

\twocolumn[
\begin{center}

{\LARGE\bfseries
The~two~faces~of~tides:~gravitational~instability~during~galaxy~mergers
}


{\large
Trisha Khan$^{1,2}$\orcidlink{0009-0004-6152-0664}\footnote{E-mail: \href{mailto:trisha.khan@res.christuniversity.in}{\textcolor{black}}{trisha.khan@res.christuniversity.in}}
\;and\;
Ayush Hazarika$^{3,4,2}$\orcidlink{0009-0004-5255-0730}\footnote{E-mail: \href{mailto:ayush.hazarika4work@gmail.com}{\textcolor{black}}{ayush.hazarika4work@gmail.com}}
}

\setcounter{footnote}{0}
\renewcommand{\thefootnote}{\arabic{footnote}}

\vspace{0.3cm}

{\small

$^{1}$\textit{Department of Physics and Electronics, CHRIST (Deemed to be University), Bengaluru 560029, India}

$^{2}$\textit{Indian Institute of Astrophysics, II Block, Koramangala, Bengaluru 560034, India}

$^{3}$\textit{International~Center~for~Theoretical~Physics~Asia-Pacific~(ICTP-AP),~University~of~Chinese~Academy~of~Sciences,~Beijing~100190,~China}

$^{4}$\textit{Taiji Laboratory for Gravitational Wave Universe, University of Chinese Academy of Sciences, Beijing 100049, China}

}

\thispagestyle{empty}

\end{center}


\begin{abstract}

Gravitational instability is a fundamental mechanism driving collapse of interstellar gas, yet in dynamically evolving environments such as galaxy mergers, external tidal fields can significantly alter the conditions for collapse. In this work, we develop an analytical framework to investigate how the tidal field of a companion galaxy modifies the classical Jeans instability by incorporating its anisotropic and time-dependent nature into the dispersion relation. We find that the tidal field introduces a critical angle between the cloud's position vector and the merger axis, $\theta_c \approx 54.7^\circ$, separating disruptive and compressive regimes of the radial tidal component. Disruptive tides suppress instability by increasing the characteristic length scale and restricting the range of unstable modes, whereas compressive tides enhance collapse by extending the unstable spectrum and increasing the growth rate of perturbations. The impact of tidal fields depends sensitively on gas density, being significant in diffuse media but negligible in dense molecular environments where the classical Jeans limit is recovered. All tidal effects peak near pericentric passage due to the strong dependence on galaxy separation. Since the free-fall time of diffuse gas is comparable to the duration of the close passage, tidally assisted collapse in the diffuse medium lags pericentre by roughly a free-fall time, whereas dense gas responds essentially instantaneously. These results demonstrate that companion-induced tidal fields play a key role in regulating the scale and efficiency of gravitational instability of diffuse interstellar gas in galaxy mergers. The code underlying this work is publicly available.

\end{abstract}

\vspace{0.3cm}

\vspace{0.5cm}

]

\footnotetext[1]{E-mail: \href{mailto:trisha.khan@res.christuniversity.in}{trisha.khan@res.christuniversity.in}}

\footnotetext[2]{E-mail: \href{mailto:ayush.hazarika4work@gmail.com}{ayush.hazarika4work@gmail.com}}

\section{Introduction}

Gravitational instability governs the collapse of gas and the formation of structure across a wide range of astrophysical environments \citep{jeans1902stability, mckee2007theory}. The classical Jeans criterion provides the fundamental condition for collapse in an isolated medium, where the competition between pressure and self-gravity determines the characteristic mass and length scales \citep{jeans1902stability}. However, astrophysical systems are rarely isolated. Gas clouds are embedded within complex and evolving gravitational potentials, where external tidal fields can significantly modify the conditions for instability \citep{Jog:2013yza}.

The role of external tides in gravitational collapse has been investigated in several analytical and observational studies. \citet{Jog:2013yza} demonstrated that tidal fields enter the dispersion relation as an additional term, modifying the effective Jeans length depending on whether the tides are compressive or disruptive. This framework has been widely adopted to study fragmentation in molecular clouds, including applications to star-forming regions \citep{zavala2023effect}. More recently, \citet{li2024modification} revisited this problem and showed that the anisotropic nature of tidal fields fundamentally alters the collapse process. In particular, it was demonstrated that collapse can proceed differently along different spatial directions, leading naturally to anisotropic fragmentation and filament formation. The importance of tides is further emphasized by the result that they can dominate the collapse dynamics over a large fraction of cloud volume.

Despite these important advances, current theoretical treatments of tidal-modified Jeans instability remain fundamentally limited in one crucial aspect: the physical origin and dynamical evolution of the tidal field are not treated in a self-consistent astrophysical context \citep{Jog:2013yza, zavala2023effect, li2024modification}. In particular, existing studies typically assume static tidal fields or highly idealized configurations (e.g. uniform or spherically symmetric external potentials), thereby neglecting the intrinsically time-dependent and geometrically structured nature of tidal interactions in realistic astrophysical systems.

This limitation becomes severe in the context of galaxy interactions and mergers, where the tidal field is set by the gravitational interaction between the galaxies and evolves dynamically along the orbit \citep{barnes1992dynamics, springel2005modelling}. In such systems, the tidal field is neither static nor isotropic: it evolves rapidly along the orbit, reaches extreme amplitudes near pericentric passage, and possesses a well-defined directional structure set by the relative geometry of the interacting galaxies \citep{barnes1992dynamics, MihosHernquist1996, renaud2015parsec}. As a result, the instability of gas is expected to depend not only on the strength of the tidal field, but also on its orientation and time dependence \citep{Jog:2013yza, li2024modification, renaud2015parsec}.

Crucially, to our knowledge, no existing analytical framework explicitly connects the Jeans instability criterion to the tidal field generated by a companion galaxy in an orbital setting. In particular, the three key aspects remain insufficiently explored: (i) how the anisotropic tidal field generated by a companion galaxy influences the stability of gas in a direction-dependent manner; (ii) how the orientation of a gas perturbation relative to the merger axis affects the onset of instability; and (iii) how the time-dependent evolution of the tidal field, driven by the orbital motion (especially near pericenter), modifies the stability criterion and growth of perturbations.

These omissions are non-trivial. As highlighted by \citet{li2024modification}, tides are intrinsically inhomogeneous and anisotropic, and their effects cannot be captured by a single scalar modification to the Jeans mass. However, even this improved anisotropic framework does not incorporate the dynamical origin of the tidal field, which is essential in merger-driven environments where the tidal tensor evolves on timescales comparable to the collapse time of the diffuse gas (see \ref{app:timescales}).

In this work, we address this gap by developing an analytical description of gravitational instability in the presence of the tidal field generated by a companion galaxy. By expanding the companion potential in the limit where the cloud size is small compared to the galaxy separation, we isolate the leading-order tidal contribution and retain its angular dependence and time evolution along the orbit. This enables us to examine how the geometry and temporal variation of merger-driven tidal fields modify the instability criterion.

Our approach bridges the gap between idealized tidal Jeans analyses and physically motivated tidal environments encountered in galaxy interactions. It provides a controlled framework to study how companion-induced tides regulate gravitational collapse, offering new insight into instability dynamics and fragmentation in dynamically evolving systems.

The paper is organized as follows. In \cref{sec:Theory} we present the theoretical framework. In \cref{sec:result-discussion} we present the results and discuss the implications of our findings. Finally, we summarize our conclusions in \cref{sec:conclusion}. Supporting material is provided in three appendices: \ref{app:dispersion} derives the modified dispersion relation from the linearized fluid equations, \ref{app:tidal field} derives the companion tidal field and its evolution, and \ref{app:timescales} compares the cloud free-fall time with the duration of the companion's passage.

\section{Theoretical Framework}\label{sec:Theory}
\subsection{Gravitational Instability in the Presence of an External Tidal Field}
The classical Jeans instability describes the collapse of a self-gravitating fluid under its own gravity \citep{jeans1902stability}. In the absence of external gravitational perturbations, the dispersion relation for density perturbations in a homogeneous medium is given by
\begin{equation}
\omega^2 = c_s^2 k^2 - 4\pi G \rho_0 ,
\end{equation}
where $c_s$ is the sound speed of the gas, $k$ is the perturbation wavenumber, and $\rho_0$ is the unperturbed background density.
\subsubsection{Modified dispersion relation}
When the cloud is embedded in an external gravitational field, the curvature of the potential modifies the instability criterion.
Following \cite{Jog:2013yza}, the dispersion relation becomes
\begin{equation}
\omega^2 = c_s^2 k^2 - 4\pi G \rho_0 + T_0 .
\label{eq:dispersion}
\end{equation}

Here $T_0 \equiv -\partial^2\Phi_{\rm ext}/\partial r^2$ is the tidal force per unit displacement along the direction of wave propagation \citep{Jog:2013yza}: $T_0>0$ corresponds to a disruptive (stretching) field that opposes the self-gravity of the perturbation, while $T_0<0$ corresponds to a compressive field that reinforces it. We emphasise that \cref{eq:dispersion} is one-dimensional, it governs plane-wave perturbations with wavevector along $\hat{\mathbf r}$ and it reduces to the classical relation when $T_0 \to 0$. A derivation from the linearized fluid equations is given in \ref{app:dispersion}.

A convenient way to describe this tidal contribution is through the tidal tensor, defined as
\begin{equation}
\tau_{ij} =
- \frac{\partial^2 \Phi_{\rm ext}}
{\partial x_i \partial x_j},
\label{eq:tidaltensor}
\end{equation}

where $\Phi_{\rm ext}$ is the external gravitational potential and
$x_i,x_j$ denote the components of an orthogonal coordinate system.With this sign convention, a positive (negative) diagonal component of $\tau_{ij}$ corresponds to a tidal force that stretches (compresses) matter along the corresponding axis.

For a spherically symmetric gravitational potential $\Phi(r)$ the tidal tensor becomes diagonal and can be written as
\begin{equation}
\tau =
\begin{pmatrix}
-\dfrac{\partial^2 \Phi}{\partial r^2} & 0 & 0 \\
0 & -\dfrac{1}{r}\dfrac{\partial \Phi}{\partial r} & 0 \\
0 & 0 & -\dfrac{1}{r}\dfrac{\partial \Phi}{\partial r}
\end{pmatrix}.
\label{eq:diagonal_comp}
\end{equation}
The first component corresponds to the radial tidal field, while the remaining two components represent the tangential tidal components.

\subsubsection{Jog Length and Mass}
By performing a linear perturbation analysis of a gas cloud embedded in an external gravitational potential, \cite{Jog:2013yza} showed that the classical Jeans instability criterion is modified by the presence of tidal forces. The characteristic length and mass scales for gravitational instability are therefore altered. We denote these modified quantities as the Jog length and Jog mass, $\lambda_{\rm Jog}$ and $M_{\rm Jog}$ (the `effective Jeans' scales of \citealt{Jog:2013yza}).

The modified instability scale can be written in terms of the classical Jeans length $\lambda_J$ as
\begin{equation}
\lambda_{\rm Jog}
=\frac{\lambda_J}{\sqrt{1-\rho_{\rm eff}/\rho_0}}, \label{eq:jog-jeans-length}
\end{equation}
and the corresponding Jog mass becomes
\begin{equation}
M_{\rm Jog}
=\frac{M_J}{\left(1-\rho_{\rm eff}/\rho_0\right)^{3/2}} . \label{eq:jog-jeans-mass}
\end{equation}
Here $\rho_0$ is the unperturbed density of the medium whose stability is being evaluated (the gas cloud). The quantity $\rho_{\rm eff}$ is an effective density introduced by the external tidal field and is defined as
\begin{equation}
\rho_{\rm eff} =
\frac{T_0}{4\pi G}.\label{eq:eff-density}
\end{equation}
The sign of $T_0$ determines the nature of the tidal field. For $T_0>0$ the tidal field is disruptive and increases the critical length and mass required for collapse. Conversely, for $T_0<0$ the tidal field is compressive and lowers the threshold for gravitational
instability.

Note that $\rho_{\rm eff}$ as defined here carries the sign of $T_0$ (\citealt{Jog:2013yza} defines it with $|T_0|$ for the compressive case): $\rho_{\rm eff}>0$ for disruptive fields and $\rho_{\rm eff}<0$ for compressive ones, so that \cref{eq:jog-jeans-length} and \cref{eq:jog-jeans-mass} apply uniformly to both regimes. These expressions are meaningful for $\rho_{\rm eff}<\rho_0$; for $\rho_{\rm eff}\ge\rho_0$ (equivalently $T_0 \ge 4\pi G \rho_0$) the right-hand side of \cref{eq:dispersion} is positive for all $k$, no scale is unstable, and the medium is tidally stabilised.

\subsection{Tidal Field During Galaxy Mergers}\label{sec:tidal field during merger}

\begin{figure}
    \centering
    \includegraphics[width=1\linewidth]{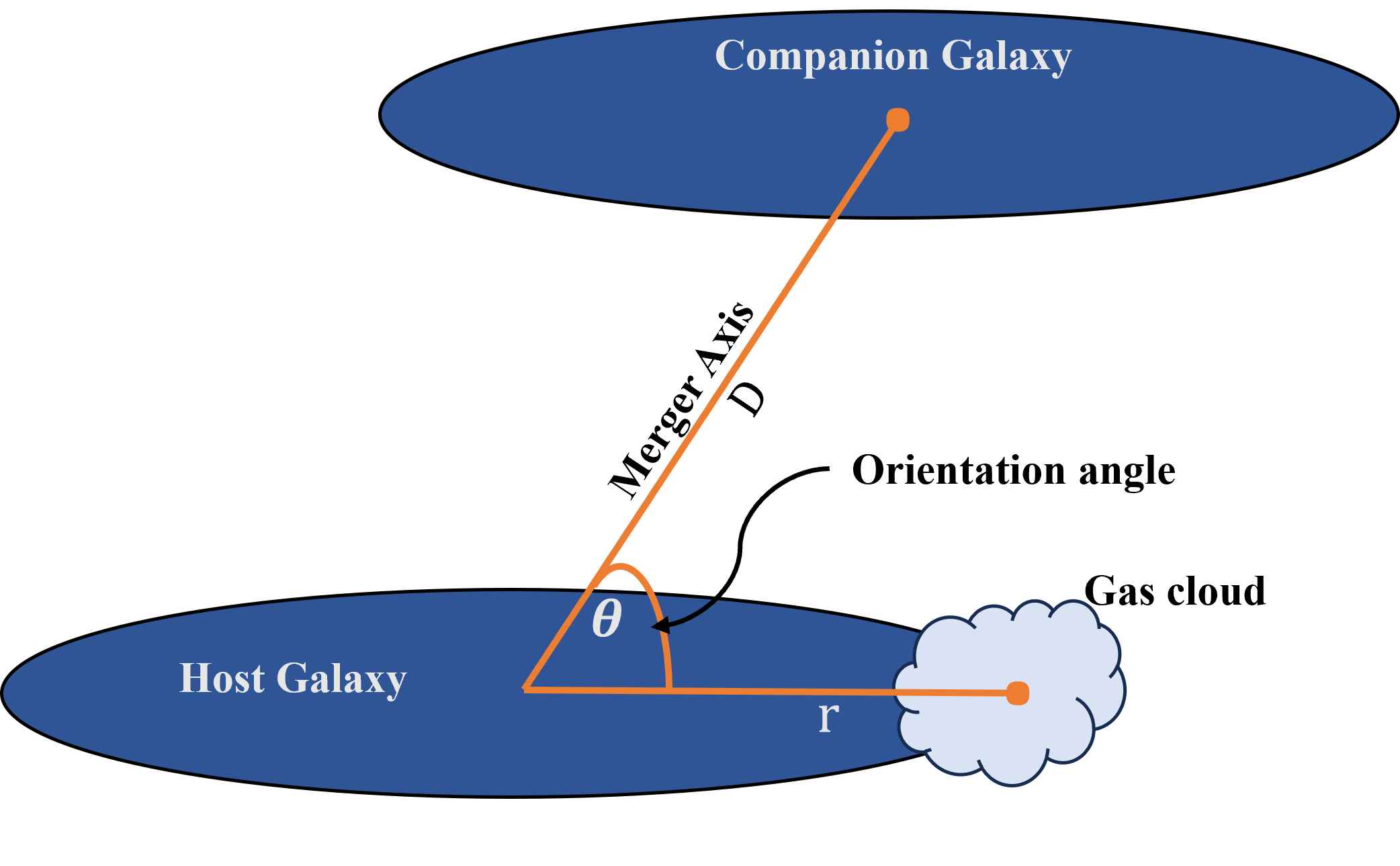}
    \caption{Schematic representation of the system considered in this work (not to scale). A gas cloud embedded in the host galaxy at position $\mathbf{r}$ experiences the tidal field of a companion galaxy located at separation $\mathbf{D}$. The angle $\theta$ defines the orientation of the cloud with respect to the merger axis, introducing a directional dependence in the tidal field.}
    \label{fig:Schematic}
\end{figure}

We consider a gas cloud embedded within the host galaxy in the presence of a companion galaxy located at a separation $\mathbf{D}(t)$ from the host centre, where $D(t)=|\mathbf{D}(t)|$. The position of the cloud relative to the host centre is denoted by $\mathbf r$ (see \cref{fig:Schematic}). The distance of the cloud from the companion galaxy is therefore
\begin{equation}
R_c = |\mathbf r - \mathbf D(t)|. \label{eq:Rc}
\end{equation}

Throughout this paper, ``radial'' refers to the direction $\hat{\mathbf r}$ of the cloud's position vector measured from the host centre. It should not be confused with the direction toward the companion, with which it coincides only for $\theta = 0$.

The gravitational potential generated by the companion galaxy is then
\begin{equation}
\Phi_{\rm comp}(R_c) = -\frac{G M_c}{R_c}. \label{eq:comp-pot}
\end{equation}
Here the companion is treated as a point mass; its finite extent softens the tide at small separations and generates structures such as extended tidal tails, which we do not model.
Since the gas cloud is embedded within the host galaxy, its distance from the host center is typically much smaller than the separation between the two galaxies, i.e. $r \ll D$. Under this condition the potential of the companion galaxy can be expanded about the host galaxy center using a Taylor expansion. Retaining terms up to second order yields
\begin{equation}
\begin{split}
\Phi_{\rm comp}(\mathbf r) \approx
 -\frac{G M_c}{D}
& -\frac{G M_c}{D^3}(\mathbf r\cdot \mathbf D)\\
&-\frac{G M_c}{2D^5}
\left[3(\mathbf r\cdot \mathbf D)^2-r^2D^2\right]. \label{eq:comp-pot-big}
\end{split}
\end{equation}
A detailed derivation is provided in \ref{App: taylor expansion}. The first term is constant and therefore does not contribute to the tidal field, while the second term corresponds to a uniform acceleration acting on the host galaxy and can be absorbed into the reference frame. The third term represents the tidal potential responsible for the differential gravitational forces acting on the gas cloud.

Introducing the angle $\theta$ between the cloud position vector and the galaxy--galaxy axis (see \cref{fig:Schematic}),
\begin{equation}
\cos\theta = \frac{\mathbf r\cdot \mathbf D}{rD},
\end{equation}
the tidal part of the potential can be written as
\begin{equation}
\Phi_{\rm tidal} =
-\frac{G M_c}{2D^3} r^2 (3\cos^2\theta - 1). \label{eq:dd-tidal}
\end{equation}
Using the definition of the tidal tensor given in
\cref{eq:tidaltensor}, the radial component of the tidal field produced by the companion galaxy becomes
\begin{equation}
\tau^{\rm comp}_{rr} =
\frac{G M_c}{D^3}(3\cos^2\theta - 1).\label{eq:tidal-radial-comp}
\end{equation}
The expansion retains the leading (quadrupole) tidal term; the neglected octupole contribution is smaller by a factor of order $r/D$, i.e. $\lesssim 25$ per cent for clouds within $r \lesssim 3$~kpc of the host centre at $D \ge r_p = 12$~kpc, with the approximation degrading for clouds in the outer disc near pericentre.
\subsubsection{Orientation Dependence of the Companion Tidal Field}
The radial tidal field generated by the companion galaxy depends on the angle $\theta$ between the position vector of the gas cloud and the galaxy--galaxy separation vector. Using the result \cref{eq:tidal-radial-comp}, we can examine several important limiting cases. 

\textit{Case I: Cloud located along the merger axis} $(\theta = 0)$. 

In this configuration the gas cloud lies along the line connecting the centers of the two galaxies. Substituting $\cos\theta = 1$ gives
\begin{equation}
\tau^{\rm comp}_{rr} = \frac{2GM_c}{D^3}.
\end{equation}
The tidal field is positive and therefore disruptive. In this case the tidal force stretches the gas cloud along the radial direction and tends to suppress gravitational collapse. Physically, the side of the cloud nearer the companion is pulled more strongly than the far side, and the differential force stretches the cloud along $\hat{\mathbf r}$.

\textit{Case II: Cloud located perpendicular to the merger axis} $(\theta = \pi/2)$

If the gas cloud lies in a direction perpendicular to the line joining the two galaxies, then $\cos\theta = 0$ and
\begin{equation}
\tau^{\rm comp}_{rr} = -\frac{GM_c}{D^3}.
\end{equation}

The tidal field is now negative: the field is compressive \emph{along} $\hat{\mathbf r}$ (while remaining disruptive along the merger axis itself). Physically, for material displaced perpendicular to the merger axis, the companion's attraction has a component directed back toward that axis which grows linearly with $r$ a transverse focusing that compresses the cloud along its radial direction. Such a configuration enhances gravitational instability along $\hat{\mathbf r}$ and promotes the collapse of gas clouds.

\textit{Case III: Critical orientation}

The tidal field changes sign when
\begin{equation}
3\cos^2\theta - 1 = 0.
\end{equation}
This condition yields a critical angle
\begin{equation}
\theta_c = \cos^{-1}\left(\frac{1}{\sqrt{3}}\right) \approx 54.7^\circ.
\end{equation}
For $\theta < \theta_c$ the tidal field is disruptive, while for $\theta > \theta_c$ it becomes compressive. Consequently, during a galaxy merger, different regions within the host galaxy may experience either tidal suppression or tidal enhancement of gravitational
instability depending on their orientation relative to the merger axis. By the $\theta \to 180^\circ-\theta$ symmetry of \cref{eq:tidal-radial-comp}, the compressive range is $54.7^\circ < \theta < 125.3^\circ$; the figures in \cref{sec:result-discussion} are therefore restricted to $\theta \in [0^\circ, 90^\circ]$ without loss of generality (\cref{fig:tcomp-time-theta-plot} shows the full range).

We emphasise that the companion's tidal tensor is trace-free, with eigenvalues $(+2GM_c/D^3,\,-GM_c/D^3,\,-GM_c/D^3)$ along and perpendicular to the merger axis (see \cref{eq:full-tensor}): the field always stretches along $\hat{\mathbf D}$ while compressing in the plane perpendicular to it, and is never compressive in all three directions. The terms ``disruptive'' and ``compressive'' in this paper therefore refer to the sign of the single component $\tau_{rr}$ entering the one-dimensional dispersion relation, tidal extension or compression along the direction of the perturbation and not to the character of the full tensor. Fully compressive tidal fields, with all three eigenvalues negative, require an extended mass distribution with a shallow density profile \citep{Das:1999fx, Renaud:2009rx} and lie outside the point-mass approximation adopted here.

\subsubsection{Time Evolution of the Galaxy Separation}

During a merger the separation between the host and companion galaxies evolves with time as the two systems orbit their common center of mass. Cosmological simulations show that galaxy encounters typically occur on highly eccentric or nearly parabolic orbits \citep{Khochfar:2003uc}. Therefore, as a first approximation, we model the relative motion of the two galaxies as a parabolic encounter.

Let $r_p$ denote the pericentric distance, i.e., the minimum separation between the two galaxies. For a parabolic orbit the instantaneous separation between the galaxies can be written as
\begin{equation}
D(t) = \frac{2 r_p}{1+\cos f(t)}, \label{eq:parabolic-separation}
\end{equation}
where $f(t)$ is the true anomaly describing the orbital phase of the encounter \citep{Murray1999}. The evolution of the true anomaly with time is given by Barker's equation
\begin{equation}
t - t_p =
\sqrt{\frac{2 r_p^3}{G M_{\rm tot}}}
\left[
\tan\left(\frac{f}{2}\right)
+ \frac{1}{3}\tan^3\left(\frac{f}{2}\right)
\right],
\label{eq:barker}
\end{equation}

where $t_p$ denotes the time of pericenter passage and
$M_{\rm tot} = M_{\rm host} + M_{\rm comp}$ is the total mass of the interacting
galaxies \citep{GalacticDynamics-book, Danby1988}.

In writing \cref{eq:parabolic-separation} and \cref{eq:barker} we treat the two galaxies as point masses on a fixed two-body orbit, as adopted in previous merger studies \citep[e.g.][]{Di_Matteo_2007}; dynamical friction and tidal mass exchange, which cause real merger orbits to decay toward coalescence after a few passages, are neglected. The present description therefore applies to the first pericentric passage.

Since the tidal field produced by the companion scales as $D(t)^{-3}$, the strength of the tidal interaction increases rapidly as the galaxies approach pericenter. Substituting the time-dependent separation into the radial tidal field \cref{eq:tidal-radial-comp} gives
\begin{equation}
\tau^{\rm comp}_{rr}(t) =
\frac{GM_c}{r_p^3}
\cos^6\left(\frac{f(t)}{2}\right)
\left(3\cos^2\theta - 1\right).\label{eq:comp-rr}
\end{equation}
A detailed derivation is presented in \ref{app: full_tidal}. This expression shows that the tidal influence of the companion galaxy depends on both the instantaneous separation between the galaxies and the orientation of the gas cloud relative to the merger axis. The tidal field therefore reaches its maximum strength near pericenter ($D \approx r_p$), where tidal interactions are expected to have the strongest impact on the gravitational stability of gas clouds within the host galaxy.
\paragraph{Pericentric Passage}
The tidal interaction between the two galaxies reaches its maximum strength at pericenter, where the separation between the galaxies becomes $D=r_p$. At this stage the true anomaly satisfies $f=0$, and therefore
\begin{equation}
\cos^6\left(\frac{f}{2}\right) = 1 .
\end{equation}
Substituting this condition into \cref{eq:comp-rr} yields
\begin{equation}
\tau^{\rm comp}_{rr} =
\frac{G M_c}{r_p^3}
\left(3\cos^2\theta - 1\right).
\end{equation}
This result shows that the strength of the tidal field during the closest approach of the galaxies depends primarily on the pericentric distance $r_p$ and the orientation of the gas cloud relative to the merger axis. In particular, the tidal field scales as $r_p^{-3}$, indicating that even modest decreases in the pericentric distance can significantly enhance the tidal interaction.

For clouds located along the merger axis ($\theta=0$), the tidal field becomes
\begin{equation}
\tau^{\rm comp}_{rr} = \frac{2GM_c}{r_p^3},
\end{equation}
corresponding to a disruptive tidal field that tends to stretch the gas cloud. Conversely, for clouds located perpendicular to the merger axis ($\theta=\pi/2$), the tidal field becomes
\begin{equation}
\tau^{\rm comp}_{rr} = -\frac{GM_c}{r_p^3},
\end{equation}
which corresponds to a compressive tidal environment capable of enhancing gravitational instability along $\hat{\mathbf r}$.
\subsection{Tidal Contribution to the Instability Criterion}\label{sec:tidal-instab} 

The radial tidal field acting on a gas cloud during a galaxy merger is governed by the gravitational influence of the companion galaxy. Using the expression derived in \cref{eq:comp-rr}, the time-dependent radial tidal field represents the tidal contribution entering the modified dispersion relation \cref{eq:dispersion}. We identify the tidal contribution entering the dispersion relation as 
\begin{equation}
T_0(t) \equiv \tau^{\rm comp}_{rr}(t).
\end{equation}
The sign of $T_0(t)$ determines whether the tidal field suppresses or enhances gravitational collapse. 

Two remarks are in order. First, \cref{eq:dispersion} governs perturbations with wavevector $\mathbf k \parallel \hat{\mathbf r}$, and $\lambda_{\rm Jog}$, $M_{\rm Jog}$ characterise collapse along that direction. Since $\hat{\mathbf k}^{\rm T}\,\tau\,\hat{\mathbf k} = (GM_c/D^3)(3\cos^2\theta_k - 1)$ for any direction $\hat{\mathbf k}$ inclined at $\theta_k$ to the merger axis, the angle $\theta$ may equivalently be read as the orientation of the perturbation wavevector itself: within a single cloud, modes inclined by more than $\theta_c$ to the merger axis are tidally assisted, while modes closer to the axis are opposed; the response is intrinsically anisotropic \citep[cf.][]{li2024modification}. Second, the tensor is trace-free: the tidal components orthogonal to any chosen direction can never \emph{both} share its sign, and the isotropic average of $T_0$ vanishes. The present results are therefore directional and should not be read as an isotropic rescaling of the Jeans mass; see section~4 of \citet{Jog:2013yza} for the conditions under which the one-dimensional relation approximates the full three-dimensional response.

In terms of the effective density defined in \cref{eq:eff-density}, the instability condition becomes 
\begin{equation}
c_s^2 k^2 < 4\pi G \left(\rho_0 - \rho_{\rm eff}(t)\right).
\end{equation}

Thus, the tidal field modifies the effective self-gravity of the gas cloud. The presence of the tidal field also modifies the critical wavenumber separating stable and unstable modes. From the instability condition, the critical wavenumber can be written as
\begin{equation}
k_{\rm crit}(t) =
\sqrt{\frac{4\pi G \rho_0 - T_0(t)}{c_s^2}}. \label{eq:k_crit}
\end{equation}

Perturbations with $k < k_{\rm crit}$ are gravitationally unstable and undergo collapse, whereas modes with $k > k_{\rm crit}$ propagate as stable waves. The tidal field therefore shifts the instability threshold depending on its sign. Disruptive tidal fields ($T_0>0$) reduce $k_{\rm crit}$ and suppress collapse, while compressive tidal fields ($T_0<0$) increase $k_{\rm crit}$, allowing instability over a broader range of wavenumbers. For $T_0 > 4\pi G \rho_0$ the argument of the square root in \cref{eq:k_crit} is negative, no unstable modes exist, and the cloud is tidally stabilised at all scales; for the WNM parameters of \cref{tab:cloud-properties} at pericentre, $T_0(0^\circ)/4\pi G\rho_0 \simeq 0.48$, so this limit is approached but not reached.

Consequently, the stability of gas clouds during galaxy mergers is governed by two key factors: the time-dependent separation between the galaxies, encoded in $D(t)$, and the orientation of the cloud relative to the merger axis through the angle $\theta$.

Finally, the cloud is of course also subject to the tidal field of the host galaxy itself (\cref{eq:diagonal_comp}). At the solar radius of a Milky-Way-like host this amounts to $T_0 = 4A(A-B) \approx 1.6\times10^{-3}~{\rm Myr^{-2}}$ in terms of the Oort constants \citep{Jog:2013yza}, roughly three times the peak companion term at pericentre for our fiducial parameters ($2GM_c/r_p^3 \simeq 5.2\times10^{-4}~{\rm Myr^{-2}}$). The host contribution, however, is quasi-static over the encounter and sets a fixed offset in the stability threshold at a given galactocentric radius, whereas the companion term carries the time and orientation dependence that are the subject of this work; we therefore isolate the latter. A combined treatment, including cored host potentials, for which the host field itself becomes compressive \citep{Das:1999fx}, is deferred to future work.

\subsection{Wave Propagation and Growth of Perturbations}

The dispersion relation given in \cref{eq:dispersion} determines the dynamical evolution of density perturbations in the presence of the companion tidal field. With $T_0(t)$ given by \cref{eq:comp-rr}, the dispersion relation becomes fully time- and orientation-dependent. 

The behavior of perturbations depends on the sign of $\omega^2$. If
\begin{equation}
c_s^2 k^2 - 4\pi G \rho_0 + T_0(t) > 0,
\end{equation}
then $\omega$ is real and the perturbations propagate as stable oscillatory waves.

Conversely, if
\begin{equation}
c_s^2 k^2 - 4\pi G \rho_0 + T_0(t) < 0,
\end{equation}
the frequency becomes imaginary and the perturbations grow exponentially. Writing $\omega^2 = -\gamma^2$ with $\gamma>0$, perturbations grow as $\rho_1 \propto e^{\gamma t}$ with growth rate 
\begin{equation}
\gamma =
\sqrt{4\pi G \rho_0 - c_s^2 k^2 - T_0(t)}. \label{eq:growth-rate}
\end{equation}

This expression shows that the companion tidal field directly modifies the rate of gravitational collapse. Disruptive tidal fields ($T_0>0$) suppress the instability by reducing the growth rate, whereas compressive tidal fields ($T_0<0$) enhance gravitational collapse. The time-dependent tidal field can therefore either stabilize or destabilize gas clouds depending on the orbital phase and the orientation relative to the merger axis.

Since $T_0(t)$ varies on the orbital timescale, \cref{eq:dispersion} with a time-dependent tidal term is to be understood in the quasi-static sense: at each orbital phase the instantaneous $T_0$ defines the instability threshold and growth rate. This is justified when the perturbation growth time is short compared with the timescale over which $T_0$ changes appreciably; \ref{app:timescales} shows that this holds comfortably for the dense phases ($t_{\rm ff} \ll t_{\rm passage}$) and only marginally for the WNM ($t_{\rm ff} \approx 0.8\,t_{\rm passage}$), for which the quoted thresholds should be read as instantaneous criteria.

\section{Results and Discussion}\label{sec:result-discussion} 
For the analysis, we adopt physically motivated parameters representative of typical galaxy merger environments. The companion galaxy mass is taken to be $M_c = 10^{11} M_\odot$ \citep{Di_Matteo_2007, GalacticDynamics-book, MihosHernquist1996}, i.e. a $\sim$1:10 companion to a Milky-Way-mass host ($M_{\rm host} = 10^{12}\,M_\odot$, as used in \ref{app:timescales}). The pericentric distance is chosen as $r_p = 12\,\mathrm{kpc}$ \citep{Di_Matteo_2007, MihosHernquist1996}, characteristic of close encounters in galaxy mergers where tidal effects are expected to be significant.

The gas density is set to $\rho_0 = 10^{-24}\,\mathrm{g\,cm^{-3}}$ \citep{McKee-WNM, Wolfire_2003, Draine2011}, corresponding to typical interstellar medium conditions. The sound speed is taken to be $c_s = 8 \times 10^5\,\mathrm{cm\,s^{-1}}$ ($\sim 8\,\mathrm{km\,s^{-1}}$) \citep{McKee-WNM, Wolfire_2003, Draine2011}, appropriate for the Warm Neutral Medium (WNM). The adopted $c_s$ should be read as an effective sound speed including a modest turbulent contribution, the purely thermal value at $T \simeq 5000$~K being $\simeq 5.6~{\rm km\,s^{-1}}$; we use these round fiducial values for the orbital-phase and dispersion figures, while the tabulated thresholds in \cref{tab:cloud-properties} are computed with the phase-specific values listed there. These choices allow us to explore the impact of merger-driven tidal fields on gravitational instability under realistic astrophysical conditions.

To examine the dependence of the instability threshold on the physical properties of the interstellar medium, we compute the Jeans wavenumber and the corresponding critical wavenumber for different gas phases. The adopted values of number density and temperature are taken from standard ISM phase classifications \citep{Draine2011}. The critical wavenumber is evaluated at pericentric passage ($f=0^\circ$), where the tidal field is strongest, for two representative orientation angles, $\theta = 0^\circ$ and $\theta = 90^\circ$, corresponding to disruptive and compressive tidal regimes, respectively. The resulting quantities are summarized in \cref{tab:cloud-properties}.

\begin{table*}
\renewcommand{\arraystretch}{1.4}
\setlength{\tabcolsep}{3pt}
\centering
\begin{tabular}{lccccccccc} 
\toprule[1.5pt]
\multirow{2}{*}{\textbf{Cloud/Phase}} 
& \textbf{$n_H$} & \textbf{$T$} & \textbf{$\rho_{0}$} & \textbf{$c_s$} & \textbf{$k_J$} & \textbf{$k_{\rm crit}(0^\circ)$} & \textbf{$k_{\rm crit}(90^\circ)$} & \textbf{$t_{\rm ff}$} & \multirow{2}{*}{\textbf{$\frac{T_0(0^\circ)}{4\pi G\rho_0}$}} \\
[-0.6em]
& \textbf{(cm$^{-3}$)} & \textbf{(K)} & \textbf{(g cm$^{-3}$)} & \textbf{(cm s$^{-1}$)} & \textbf{(cm$^{-1}$)} & \textbf{(cm$^{-1}$)} & \textbf{(cm$^{-1}$)} & \textbf{(Myr)} & \\
\midrule
WNM & 0.6 & 5000 & $1.31\times10^{-24}$ & $5.63\times10^{5}$ & $1.86\times10^{-21}$ & $1.34\times10^{-21}$ & $2.07\times10^{-21}$ & $58.26$ & $0.48$ \\
CNM & 30 & 100 & $6.53\times10^{-23}$ & $7.97\times10^{4}$ & $9.29\times10^{-20}$ & $9.24\times10^{-20}$ & $9.31\times10^{-20}$ & $8.24$ & $9.6\times10^{-3}$ \\
Diffuse H$_2$ & 100 & 50 & $2.18\times10^{-22}$ & $5.63\times10^{4}$ & $2.40\times10^{-19}$ & $2.39\times10^{-19}$ & $2.40\times10^{-19}$ & $4.51$ & $2.9\times10^{-3}$ \\
Dense H$_2$ & $10^4$ & 30 & $2.18\times10^{-20}$ & $4.36\times10^{4}$ & $3.10\times10^{-18}$ & $3.10\times10^{-18}$ & $3.10\times10^{-18}$ & $0.45$ & $2.9\times10^{-5}$ \\
GMC & $10^3$ & 15 & $2.18\times10^{-21}$ & $3.09\times10^{4}$ & $1.38\times10^{-18}$ & $1.38\times10^{-18}$ & $1.38\times10^{-18}$ & $1.43$ & $2.9\times10^{-4}$ \\
\bottomrule[1.5pt]
\end{tabular}
\caption{Cloud properties, free-fall timescale ($t_{\rm ff}$), classical Jeans wavenumber ($k_J$), and tidal-field-modified critical wavenumber ($k_{\rm crit}$) for representative interstellar medium phases. All quantities are evaluated at pericenter ($f=0^\circ$), for orientation angles $\theta=0^\circ$ and $\theta=90^\circ$. The final column lists the ratio of the disruptive tidal term at pericentre to the self-gravity term (the compressive case, $\theta=90^\circ$, has half this magnitude), which controls the deviation of $k_{\rm crit}$ from $k_J$. Sound speeds are thermal values for the adopted temperatures with $\mu \simeq 1.3$; for the H$_2$-dominated phases a molecular mean weight ($\mu \simeq 2.33$) would give lower $c_s$ and an even smaller relative tidal correction.}\label{tab:cloud-properties}
\end{table*}

In addition to the Jeans wavenumber and the tidal-field-modified critical wavenumber, we compute the gravitational free-fall timescale of each cloud, defined in \cref{eq:free-fall time}. The free-fall timescale represents the characteristic time required for a pressure-free, self-gravitating cloud to collapse under its own gravity \citep{GalacticDynamics-book}. The corresponding values are also listed in \cref{tab:cloud-properties}.

The results presented in \cref{tab:cloud-properties} reveal a clear and physically intuitive dependence of the instability scale on the properties of the interstellar medium. In particular, the comparison between the classical Jeans wavenumber $k_J$, and the tidal-field-modified critical wavenumber $k_{\rm crit}(t)$, as seen in \cref{eq:k_crit} highlights the relative importance of self-gravity and the external tidal field.

The key quantity governing the deviation from the Jeans criterion is the ratio of the tidal term to the self-gravity term, i.e. $T_0/(4\pi G\rho_{0})$ (final column of \cref{tab:cloud-properties}). In diffuse phases such as the WNM, the gas density is low, and consequently the self-gravitational term $4\pi G\rho_0$ is small. In this regime, the tidal contribution becomes comparable in magnitude, leading to a noticeable difference between $k_{\rm crit}$ and $k_J$. As seen in \cref{tab:cloud-properties}, this results in a clear separation between the two quantities, with the direction of the shift determined by the sign of $T_0$ (disruptive or compressive). 

As we move toward denser and colder phases like Cold Neutral Medium (CNM), diffuse H$_2$, etc., the increase in density leads to a corresponding increase in the self-gravitational term. Although the tidal field remains unchanged for fixed orbital parameters, its relative influence diminishes, and the deviation between $k_{\rm crit}$ and $k_J$ becomes progressively smaller.

In the densest regimes, such as giant molecular clouds (GMCs) and dense H$_2$, the condition $4\pi G\rho_0 \gg |T_0|$ is satisfied. In this limit, the tidal contribution becomes negligible, and the instability criterion effectively reduces to the classical Jeans condition, i.e. $k_{\rm crit} \simeq k_J$. This behaviour is clearly reflected in \cref{tab:cloud-properties}, where the two quantities converge for the densest phases.

Physically, this trend indicates that tidal fields can significantly modify gravitational instability in diffuse gas, where self-gravity is weak, but become increasingly irrelevant in dense molecular environments where self-gravity dominates. As a result, tidal effects primarily regulate the onset of instability in low-density phases, while the subsequent fragmentation and collapse in star-forming regions proceed largely according to the classical Jeans picture.

\begin{figure}[!t]
    \centering
    \includegraphics[width=1\linewidth]{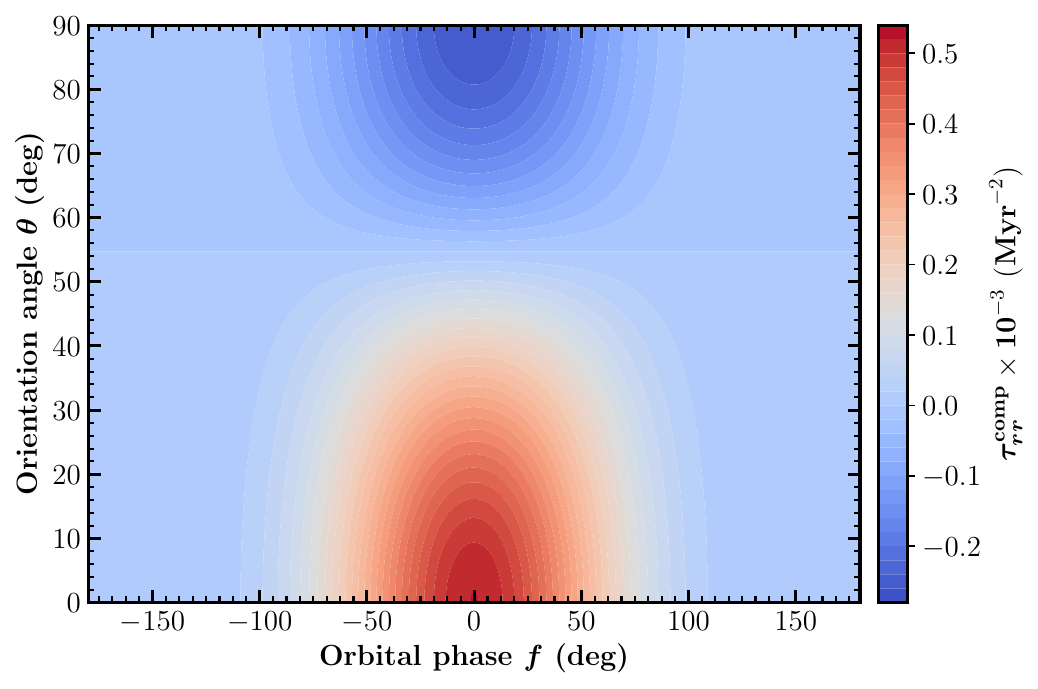}
    \caption{2-D contour map of the radial component of the companion tidal field, $\tau_{rr}^{\mathrm{comp}}$, shown as a function of orbital phase $f$ and orientation angle $\theta$. The tidal field is computed using \cref{eq:comp-rr} and plotted in units of $10^{-3}\,\ \mathrm{Myr}^{-2}$. The adopted parameters are $G = 4.498\times10^{-12}\ \mathrm{kpc}^3 \,\ \mathrm{M}_\odot^{-1}\,\ \mathrm{Myr}^{-2}$, $M_c = 10^{11}\ \mathrm{M}_\odot$, and pericentric distance $r_p = 12\ \mathrm{kpc}$. The orbital phase spans $f \in [-180^\circ, 180^\circ]$, with $f=0^\circ$ corresponding to pericenter, while $\theta \in [0^\circ, 90^\circ]$ denotes the orientation angle. Positive (red) values correspond to disruptive tidal fields, while negative (blue) values indicate compressive tidal fields.} 
    \label{fig:theta-f}
\end{figure}

\cref{fig:theta-f} shows the 2-D distribution of the radial tidal field $\tau_{rr}^{\rm comp}$ as a function of orbital phase $f$ and orientation angle $\theta$, computed using \cref{eq:comp-rr}. The structure of the tidal field directly reflects the separable dependence on orbital evolution and geometric orientation.

The overall amplitude of the tidal field is governed by the factor $\cos^6(f/2)$, which encodes the time dependence of the galaxy separation. As a result, the tidal field reaches its maximum near pericentric passage ($f=0^\circ$), where $\cos^6(f/2)=1$, and rapidly decreases as the system moves away from pericenter. This produces the strong concentration of high-magnitude tidal regions around $f=0^\circ$ seen in \cref{fig:theta-f}.

The angular dependence is controlled by the term $(3\cos^2\theta - 1)$, which determines both the sign and anisotropy of the tidal field. For $\theta < \theta_c \approx 54.7^\circ$, this term is positive, corresponding to a disruptive tidal field that suppresses gravitational collapse. In contrast, for $\theta > \theta_c$, the tidal field becomes negative, indicating a compressive regime that enhances instability. The transition between these regimes is clearly visible in the contour map as a sign change across the critical angle.

The combination of these two effects results in a highly anisotropic and time-dependent tidal field. In particular, compressive tidal regions are most prominent near pericenter and for orientations perpendicular to the merger axis, highlighting the importance of both orbital phase and geometry in regulating gravitational instability during galaxy mergers.

\begin{figure}[!t]
    \centering
    \includegraphics[width=1\linewidth]{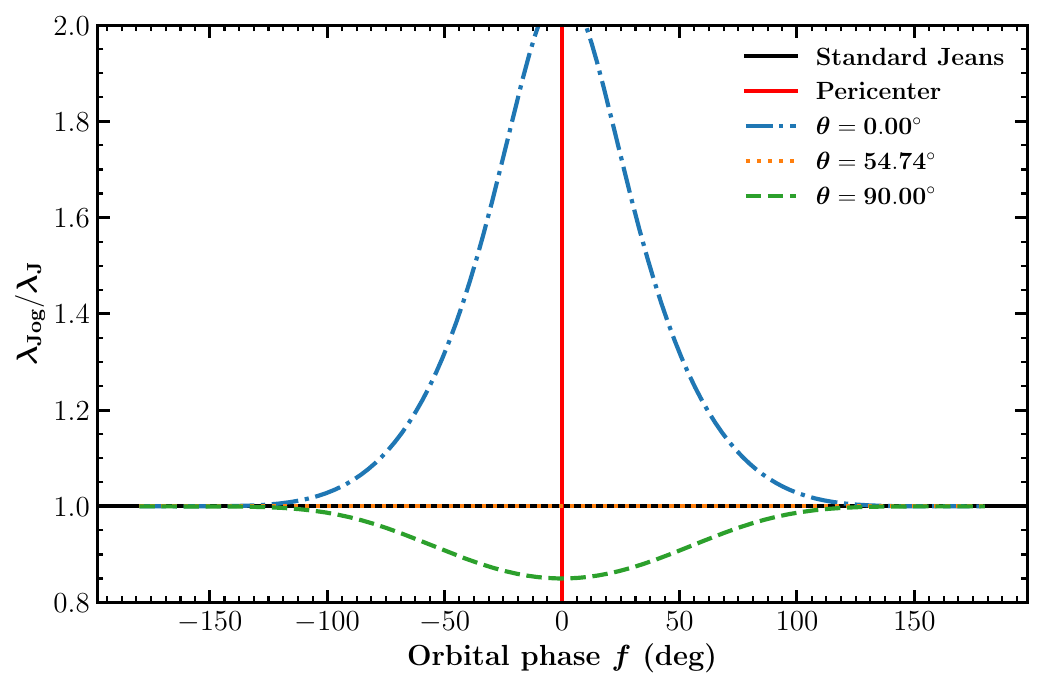}
    \caption{Ratio of the Jog length $\lambda_{\mathrm{Jog}}$, to the classical Jeans length, $\lambda_J$, plotted against orbital phase $f$ for different orientation angles $\theta$. The ratio is computed using \cref{eq:jog-jeans-length}. The adopted parameters are the same as in \cref{fig:theta-f}. In addition, the gas density value $\rho_0 = 1.2\times10^{7}\ \mathrm{M}_\odot \,\ \mathrm{kpc}^{-3}$ ($\simeq 8.1\times10^{-25}~{\rm g\,cm^{-3}}$), representative of the WNM. The orbital phase spans $f \in [-180^\circ, 180^\circ]$, with $f=0^\circ$ marking pericenter (vertical red line). The horizontal black line at unity corresponds to the classical Jeans length in the absence of tidal effects. The blue dash–dotted, orange dotted, and green dashed curves correspond to orientation angles $\theta = 0^\circ$, $54.74^\circ$, and $90^\circ$, respectively.}
    \label{fig:Ljog/Ljeans}
\end{figure}

\cref{fig:Ljog/Ljeans} shows the ratio $\lambda_{\rm Jog}/\lambda_J$ as a function of orbital phase $f$ for different orientation angles $\theta$. The variation of this ratio directly reflects the influence of the tidal field through \cref{eq:jog-jeans-length}.

A strong dependence on both orbital phase and orientation is evident. Near pericentric passage ($f=0^\circ$), the deviation from the classical Jeans length is maximized, consistent with the enhanced tidal field at minimum separation. As the system evolves away from pericenter, the ratio approaches unity, indicating that tidal effects become negligible at large separations.

The nature of the tidal field can be directly inferred from the behavior of $\lambda_{\rm Jog}$. For $\theta = 0^\circ$, we observe that $\lambda_{\rm Jog} > \lambda_J$ near pericenter. Since an increase in the critical length scale implies that larger perturbations are required for collapse, this corresponds to a disruptive tidal field ($T_0 > 0$) that suppresses gravitational instability.

At $\theta \approx 54.7^\circ$, the ratio remains equal to unity for all orbital phases, i.e. $\lambda_{\rm Jog} = \lambda_J$. This indicates that the tidal contribution vanishes ($T_0 = 0$), consistent with the theoretical condition $3\cos^2\theta - 1 = 0$.

In contrast, for $\theta = 90^\circ$, we find $\lambda_{\rm Jog} < \lambda_J$ near pericenter. A reduction in the critical length scale implies that smaller perturbations can collapse, indicating a compressive tidal field ($T_0 < 0$) that enhances gravitational instability.

These results demonstrate that the tidal field in galaxy mergers can be directly diagnosed through its impact on the instability scale, with orientation and orbital phase jointly determining whether collapse is suppressed or promoted.

\begin{figure}[!t]
    \centering
    \includegraphics[width=1\linewidth]{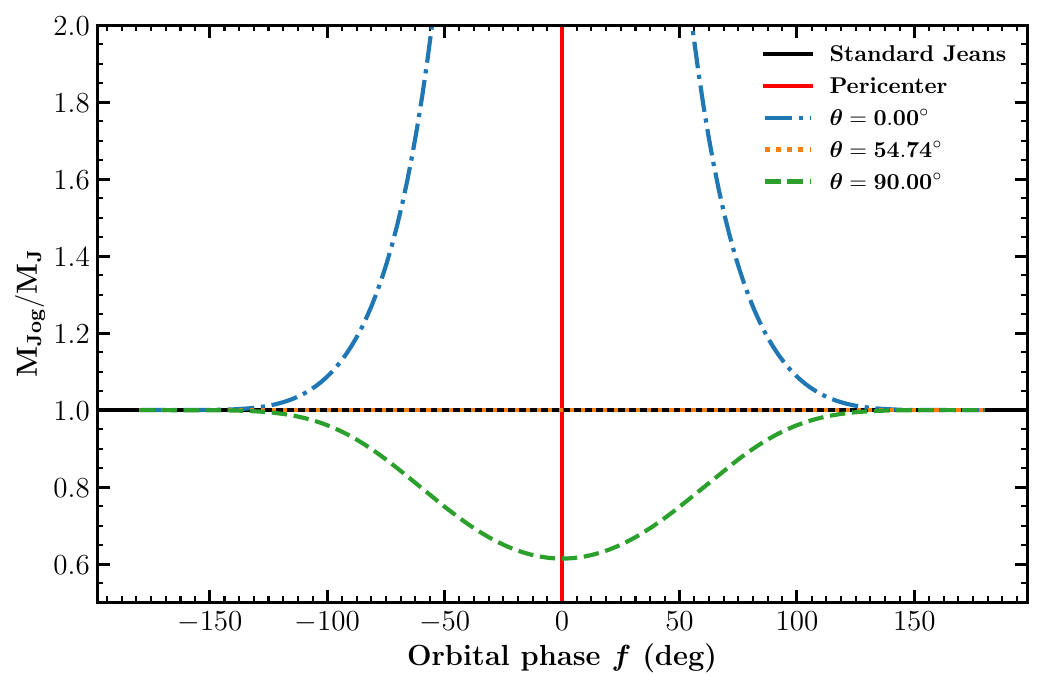}
    \caption{Same as \cref{fig:Ljog/Ljeans}, but for the ratio of the Jog mass to the classical Jeans mass, computed using \cref{eq:jog-jeans-mass}.}
    \label{fig:Mjog/Mjeans}
\end{figure}

\cref{fig:Mjog/Mjeans} shows the ratio of the Jog mass to the classical Jeans mass as a function of orbital phase for different orientations. The behavior closely follows that of the Jog length, but with a stronger quantitative response due to the cubic dependence on the instability scale (see \cref{eq:jog-jeans-mass}). In particular, the deviation from unity is significantly amplified near pericenter, indicating that tidal effects have a more pronounced impact on the mass scale of collapse than on the corresponding length scale. For orientations aligned with the merger axis ($\theta = 0^\circ$), the increase in $M_{\rm Jog}$ implies a substantial suppression of collapse, whereas for perpendicular configurations ($\theta = 90^\circ$), the reduction in $M_{\rm Jog}$ suggests that the tidal field promotes collapse on smaller mass scales. This highlights that tidal fields not only regulate whether collapse occurs but also strongly influence the characteristic mass of the resulting structures.

\begin{figure}[!t]
    \centering
    \includegraphics[width=1\linewidth]{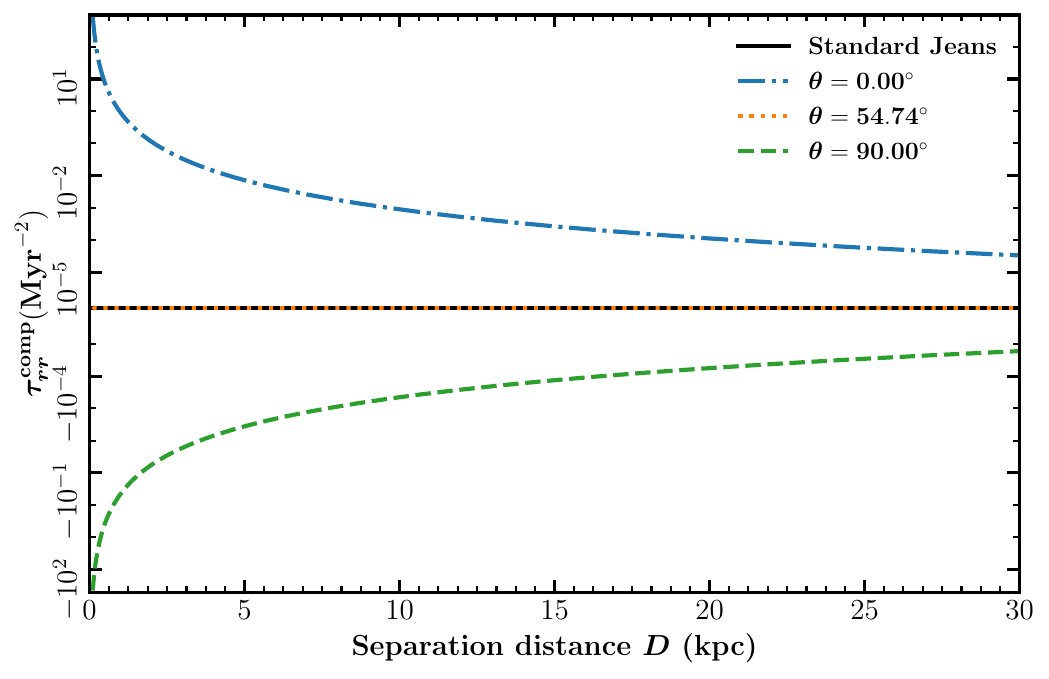}
    \caption{Variation of the radial component of the companion tidal field, $\tau_{rr}^{\rm comp}$, shown as a function of separation distance $D$ in kpc. The tidal field is computed using \cref{eq:tidal-radial-comp}, with parameters identical to those adopted in \cref{fig:theta-f}. The curves follow the same convention as in \cref{fig:Ljog/Ljeans}. The horizontal black line marks $\tau_{rr}^{\rm comp}=0$, i.e. the absence of a tidal contribution (labelled `Standard Jeans' in the legend).}
    \label{fig:D-Tcomp}
\end{figure}

\cref{fig:D-Tcomp} shows the variation of the radial tidal field $\tau^{\rm comp}_{rr}$ as a function of separation distance $D$ for different orientation angles $\theta$, computed using \cref{eq:tidal-radial-comp}. The tidal field exhibits a strong dependence on separation, decreasing rapidly with increasing $D$ due to its characteristic $D^{-3}$ scaling. This behavior highlights that tidal interactions are highly localized and become significant only during close encounters.

The sign and magnitude of the tidal field are determined by the angular factor $(3\cos^2\theta - 1)$. For $\theta = 0^\circ$, the tidal field is positive and decreases monotonically with distance, corresponding to a disruptive regime. In contrast, for $\theta = 90^\circ$, the tidal field is negative, indicating a compressive regime whose magnitude similarly weakens with increasing separation. At the critical angle $\theta \approx 54.7^\circ$, the tidal field vanishes identically for all $D$, consistent with the condition $3\cos^2\theta - 1 = 0$.

\cref{fig:D-Tcomp} reinforces that both the strength and nature of tidal interactions in galaxy mergers are controlled by the interplay between separation and orientation, with the most significant effects occurring at small $D$ and away from the critical angle.

\begin{figure}[!t]
    \centering
    \includegraphics[width=1\linewidth]{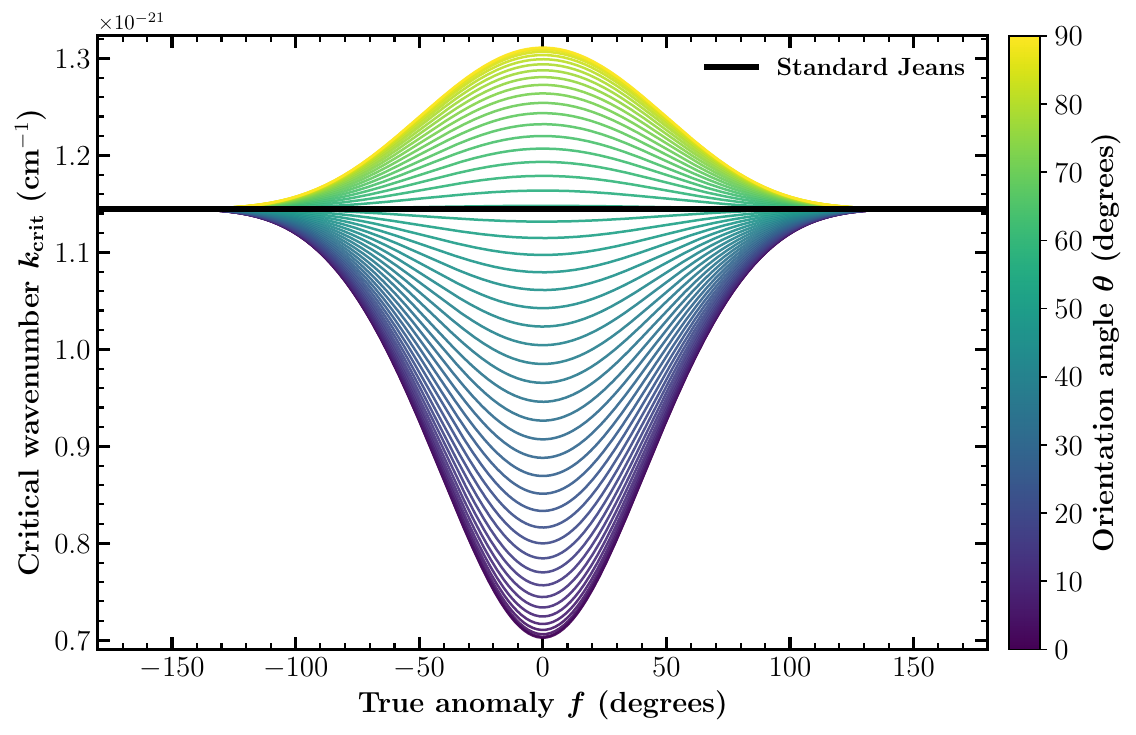}
    \caption{Variation of the critical wavenumber, $k_{\rm crit}$, as a function of true anomaly $f$ for different orientation angles $\theta$, illustrating the effect of the companion-induced tidal field on gravitational instability. The critical wavenumber is computed using \cref{eq:k_crit}. All the adopted parameters are expressed in cgs units, with $G = 6.674\times10^{-8}\ \mathrm{cm^3 \,\ g^{-1}\,\ s^{-2}}$, gas density $\rho_0 = 10^{-24}\ \mathrm{g\,\ cm^{-3}}$, and sound speed $c_s = 8\times10^{5}\ \mathrm{cm\,\ s^{-1}}$, representative of the WNM (round fiducial values; cf.\ the WNM entry of \cref{tab:cloud-properties}). The companion mass is $M_c = 10^{11}\ M_\odot$ (converted to grams), and the pericentric distance is $r_p = 12\ \mathrm{kpc}$ (converted to cm). Each curve represents a different orientation angle $\theta \in [0^\circ, 90^\circ]$, color-coded as indicated by the color bar. The horizontal black line denotes the standard Jeans wavenumber in the absence of tidal effects.}
    \label{fig:Anomaly}
\end{figure}

\cref{fig:Anomaly} shows the variation of the critical wavenumber $k_{\rm crit}$ as a function of orbital phase $f$ for different orientation angles $\theta$, computed using \cref{eq:k_crit}. The deviation of $k_{\rm crit}$ from the standard Jeans value (horizontal black line) reflects the direct influence of the tidal field on the instability threshold.

A pronounced dependence on both orbital phase and orientation is evident. Near pericentric passage ($f=0^\circ$), where the tidal field is strongest, the deviation from the classical Jeans wavenumber is maximized. As the system evolves away from pericenter, all curves converge to the Jeans value, indicating that tidal effects become negligible at large separations.

The direction of the shift in $k_{\rm crit}$ depends on the sign of the tidal contribution $T_0(t)$. For small orientation angles ($\theta \approx 0^\circ$), $k_{\rm crit}$ decreases below the Jeans value, implying that only longer-wavelength (smaller $k$) perturbations remain unstable, consistent with a disruptive tidal field that suppresses collapse. In contrast, for larger angles ($\theta \rightarrow 90^\circ$), $k_{\rm crit}$ increases above the Jeans value, allowing shorter-wavelength perturbations to become unstable, indicating a compressive tidal regime. At the critical angle $\theta \approx 54.7^\circ$, the tidal contribution vanishes and $k_{\rm crit}$ coincides with the classical Jeans value across all orbital phases.

These results demonstrate that the companion-induced tidal field shifts the instability boundary in wavenumber space, with compressive tides broadening the range of unstable modes and disruptive tides restricting it.

\begin{figure}[!t]
    \centering
    \includegraphics[width=1\linewidth]{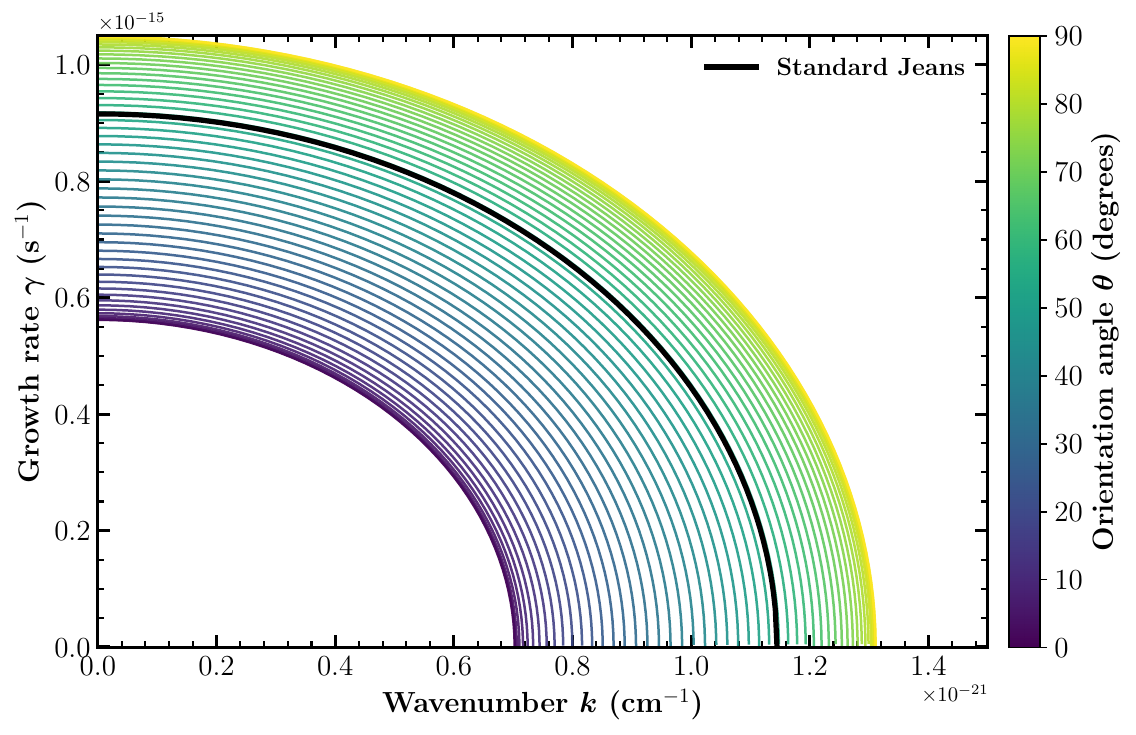}
    \caption{Variation of the growth rate $\gamma$ as a function of wavenumber $k$ at pericenter ($f = 0$), incorporating the effect of the tidal field. The growth rate is computed using \cref{eq:growth-rate}. The adopted parameters are the same as in \cref{fig:Anomaly}. The curves and color-coding also follow the same convention as in \cref{fig:Anomaly}, with the black line indicating the standard Jeans case.}
    \label{fig:growth-curve}
\end{figure}

\cref{fig:growth-curve} shows the growth rate $\gamma$ as a function of wavenumber $k$ at pericenter ($f=0$) for different orientation angles $\theta$, computed using \cref{eq:growth-rate}. The presence of the tidal field modifies both the range of unstable modes and the corresponding growth rates relative to the standard Jeans case (black curve).

A clear dependence on orientation is observed. For small angles ($\theta \approx 0^\circ$), the growth rate is suppressed across all wavenumbers, and the cutoff shifts to lower $k$, indicating that only larger-scale perturbations remain unstable. This behavior is consistent with a disruptive tidal field ($T_0>0$), which reduces the effective self-gravity of the system.

In contrast, for larger angles ($\theta \rightarrow 90^\circ$), the growth rate is enhanced and extends to higher wavenumbers, allowing smaller-scale perturbations to become unstable. This corresponds to a compressive tidal field ($T_0<0$), which effectively strengthens self-gravity and accelerates the collapse.

The maximum growth rate also increases with $\theta$, demonstrating that compressive tides not only broaden the instability range but also amplify the rate of collapse. At the critical angle $\theta \approx 54.7^\circ$, the tidal contribution vanishes and the standard Jeans behavior is recovered.

\begin{figure}[!t]
    \centering
    \includegraphics[width=1\linewidth]{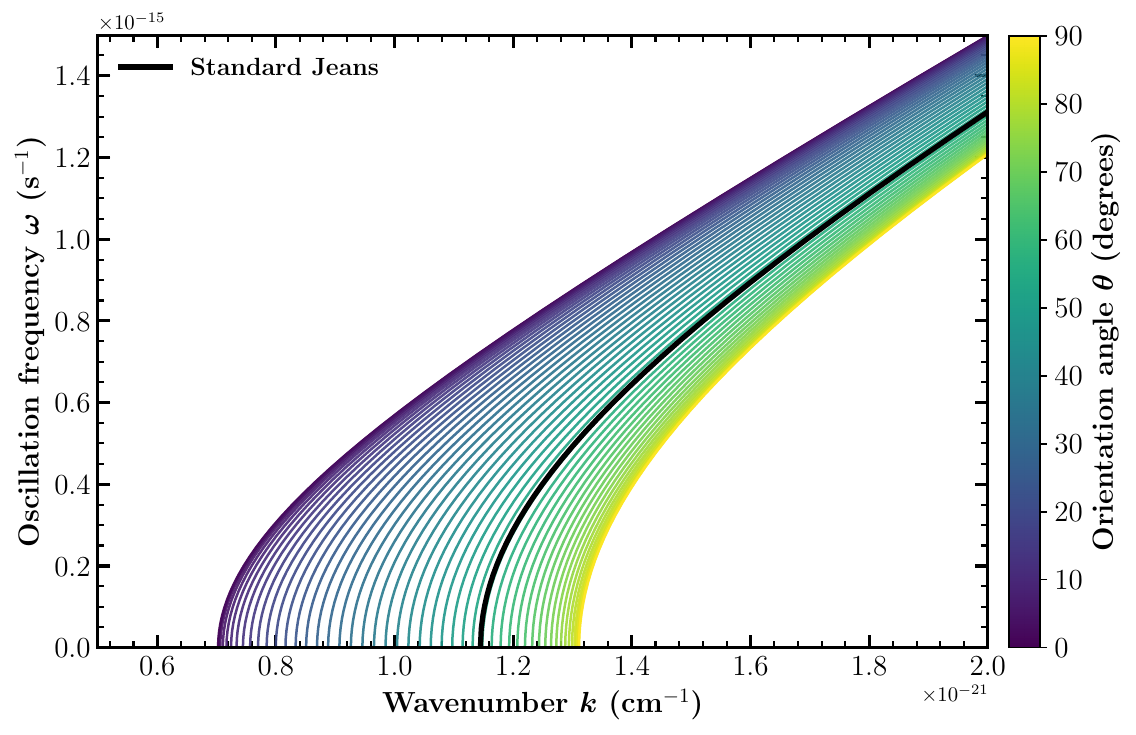}
    \caption{Same as \cref{fig:growth-curve}, but for the oscillation frequency $\omega$ as a function of wavenumber $k$ at pericenter ($f = 0$), computed using \cref{eq:dispersion}.}
    \label{fig:Oscillation-freq}
\end{figure}

These results show that tidal fields in galaxy mergers can significantly reshape both the spectrum and efficiency of gravitational instability, with compressive configurations promoting rapid collapse and disruptive configurations inhibiting it.
\cref{fig:Oscillation-freq} shows the oscillation frequency $\omega$ as a function of wavenumber $k$ at pericenter ($f=0$) for different orientation angles $\theta$, computed from the dispersion relation \cref{eq:dispersion}. This corresponds to the regime $\omega^2 > 0$, where perturbations evolve as stable oscillatory modes. A higher oscillation frequency indicates a more stable configuration, as perturbations are restored on shorter timescales compared to gravitational collapse.

The presence of the tidal field shifts the oscillatory branch relative to the standard Jeans case (black curve). For small orientation angles ($\theta \approx 0^\circ$), the disruptive field ($T_0>0$) raises $\omega^2$ at every wavenumber: the $\omega=0$ intercept moves to lower $k$ ($k_{\rm crit}<k_J$), so oscillatory behaviour sets in already at longer wavelengths, and at fixed $k$ the oscillation frequency is higher, reflecting stronger effective restoring forces. In contrast, for $\theta \rightarrow 90^\circ$ the compressive field ($T_0<0$) lowers $\omega^2$ at every wavenumber: the intercept moves to higher $k$ ($k_{\rm crit}>k_J$), so perturbations remain unstable over a broader range of wavenumbers and enter the oscillatory regime only at shorter wavelengths, with correspondingly lower frequencies near the stability boundary.

The shift in the oscillatory branch is directly linked to the modification of the critical wavenumber, with the boundary $\omega=0$ coinciding with $k_{\rm crit}$. These results demonstrate that tidal fields not only modify the growth of unstable modes but also regulate the stability of oscillatory perturbations through changes in the restoring timescale.
Overall, the results demonstrate that companion-induced tidal fields introduce a strong anisotropic and time-dependent modification to gravitational instability. Disruptive tides suppress collapse by reducing the range of unstable modes and lowering growth rates, whereas compressive tides enhance instability by extending the unstable spectrum and accelerating collapse. The interplay between orbital phase and orientation therefore plays a central role in determining both the onset and efficiency of structure formation in galaxy mergers. We note that merger simulations find compressive tidal modes and the star formation they trigger, to peak at the pericentre passages and to last a few
tens of Myr \citep{Renaud:2009rx}, with young star clusters forming almost
exclusively inside such regions \citep{renaud2015parsec}, in agreement with the timing and duration of the tidal modulation quantified here. Those fully
compressive modes, however, arise from the overlap of the extended galactic
potentials rather than from a point-mass quadrupole (\cref{sec:tidal field during merger}); the present framework isolates the time-dependent, directional ingredient of that picture and is complementary to, rather than a derivation of, those numerical results.

\section{Conclusion}\label{sec:conclusion}

In this work, we have developed an analytical framework to investigate the impact of a companion galaxy on gravitational instability in gaseous media. By incorporating the tidal field generated during a galaxy encounter, we extend the classical Jeans analysis to account for the inherently anisotropic and time-dependent nature of merger-driven tidal interactions.

Our main results can be summarized as follows:
\begin{enumerate}
    \item The tidal field generated by the companion galaxy is intrinsically anisotropic and strongly dependent on orbital phase. Its amplitude peaks near pericentric passage due to the $D^{-3}$ scaling, demonstrating that tidal effects are highly localized in time during galaxy encounters.
    \item The angular dependence of the tidal field introduces a critical orientation angle, $\theta_c \approx 54.7^\circ$, which separates disruptive ($T_0>0$) and compressive ($T_0<0$) tidal regimes. This critical angle emerges naturally from the quadrupolar structure of the tidal field and plays a central role in determining the dynamical response of the cloud. We stress that these regimes refer to the radial component of the tide: the full tensor of the companion remains of stretch, squeeze character (trace-free) at every orientation, and is never compressive in all three directions.
    \item The influence of tidal fields on gravitational instability depends sensitively on the density of the gaseous medium. In diffuse phases such as the WNM, the tidal term is comparable to the self-gravitational term, leading to significant deviations from the classical Jeans criterion. In contrast, for dense molecular gas, where $4\pi G \rho_0 \gg |T_0|$, the effect of tides becomes negligible and the instability threshold converges to the standard Jeans limit. This demonstrates that tidal fields primarily regulate instability in low-density environments, while dense star-forming regions remain largely governed by self-gravity.
    \item A comparison between the cloud free-fall timescale and the companion passage timescale shows that, for the adopted merger parameters, the tidal perturbation persists long enough to influence the dynamical evolution of interstellar clouds. In particular, the free-fall time of diffuse gas is comparable to the duration of the close encounter, whereas denser molecular clouds collapse on significantly shorter timescales. This indicates that merger-driven tidal fields are dynamically relevant throughout the interaction, while their influence becomes progressively less important once self-gravity dominates in dense clouds.
    \item Disruptive tidal fields increase the characteristic instability scale, raising both the effective Jeans length and mass. This implies that only larger-scale perturbations can collapse, leading to a suppression of gravitational instability. In contrast, compressive tidal fields reduce these scales, allowing smaller perturbations to become unstable.
    \item The instability threshold in wavenumber space is significantly modified by the tidal field. Disruptive tides reduce the critical wavenumber, thereby restricting the range of unstable modes, whereas compressive tides increase it, broadening the spectrum of perturbations that can undergo gravitational collapse.
    \item The growth rate of density perturbations is strongly influenced by the tidal environment. Disruptive tidal fields suppress the growth of perturbations by reducing the effective self-gravity, while compressive tidal fields enhance both the range and rate of unstable modes, leading to more rapid collapse.
    \item In the stable regime, the oscillation frequency of perturbations provides a direct measure of stability. Higher oscillation frequencies correspond to stronger restoring forces and enhanced stability, whereas lower frequencies indicate weaker restoring forces and proximity to the instability threshold. Tidal fields therefore modify not only the onset of instability but also the dynamical response of stable modes.
\end{enumerate}

Overall, our analysis demonstrates that companion-induced tidal fields play a fundamental role in regulating gravitational instability in galaxy mergers. The interplay between orbital evolution and geometric orientation determines both the scale and efficiency of collapse, providing a natural mechanism for controlling fragmentation in dynamically evolving systems. These results highlight the importance of incorporating realistic, time-dependent tidal fields in both analytical and numerical studies of instability dynamics in interacting galaxies.

\section*{Acknowledgements}
The authors are sincerely grateful to Mousumi Das (IIA) for her insightful comments and valuable discussions on the interpretation of compressive tidal fields, realistic astrophysical scenarios, and star formation, which significantly improved the scientific content and presentation of this manuscript. T.K. acknowledges support from the Department of Physics and Electronics, CHRIST (Deemed to be University), Bengaluru. She gratefully acknowledges Arun Roy (IIA \& CHRIST) and Blesson Mathew (CHRIST) for their constant support and encouragement. A.H. acknowledges support from the National Natural Science Foundation of China (NSFC) under Grants No. 12347103 and 12547104, as well as the Fundamental Research Funds for the Central Universities. A.H. is sincerely grateful to Andrew L. Miller (ICTP-AP) and Debika Chowdhury (IIA) for their warm hospitality and valuable support throughout the period. T.K. and A.H. further acknowledge that part of this work was carried out during their tenure as Visiting Students at the Indian Institute of Astrophysics under the Visiting Student Programme organized by the Board of Graduate Studies.

\section*{Data Availability}
This study does not use any observational or experimental data. The computational code developed and used for the analysis in this work are publicly available at \cite{codes}.

\bibliography{references}
\bibliographystyle{mnras}

\appendix
\labelformat{section}{Appendix #1}
\labelformat{subsection}{Appendix #1}
\section{Derivation of the modified dispersion relation}\label{app:dispersion}
For completeness we outline the derivation of \cref{eq:dispersion}, following \citet{Jog:2013yza}. Consider a homogeneous self-gravitating medium of density $\rho_0$ and sound speed $c_s$, subject to an external potential $\Phi_{\rm ext}$. The linearized continuity, force, and Poisson equations for the perturbations $(\rho_1, \mathbf v_1, \Phi_1)$ are
\begin{align}
\frac{\partial \rho_1}{\partial t} + \rho_0\, \nabla\cdot\mathbf v_1 &= 0, \\
\rho_0\, \frac{\partial \mathbf v_1}{\partial t} &= -\rho_0 \nabla\Phi_1 - c_s^2\, \nabla\rho_1 - \rho_1 \nabla\Phi_{\rm ext}, \\
\nabla^2 \Phi_1 &= 4\pi G \rho_1 .
\end{align}
The background force $-\rho_0\nabla\Phi_{\rm ext}$ is removed by working in the freely falling frame of the host centre, the same step by which the uniform (dipole) term of \cref{eq:comp-pot-big} was absorbed, so that the external field acts only on the perturbed density. This is the extended Jeans swindle discussed by \citet{Jog:2013yza} \citep[see also][]{GalacticDynamics-book}. Taking the time derivative of the continuity equation and the divergence of the force equation, and eliminating $\mathbf v_1$ and $\Phi_1$, gives
\begin{equation}
\frac{\partial^2 \rho_1}{\partial t^2} = c_s^2 \nabla^2\rho_1 + 4\pi G \rho_0\, \rho_1 + \nabla\cdot\!\left(\rho_1 \nabla\Phi_{\rm ext}\right).
\end{equation}
For plane-wave perturbations $\rho_1 \propto \exp[i(\omega t - kr)]$ with wavevector along $\hat{\mathbf r}$, and treating the tidal tensor as uniform across a wavelength (valid for $\lambda \ll D$; here $\lambda_J \sim 1$~kpc for the WNM, compared with $D \ge r_p = 12$~kpc), the external term reduces to $\rho_1\, \partial^2\Phi_{\rm ext}/\partial r^2 = -\tau_{rr}\, \rho_1$, and we recover
\begin{equation}
\omega^2 = c_s^2 k^2 - 4\pi G \rho_0 + T_0, \qquad T_0 \equiv \tau_{rr} = -\frac{\partial^2 \Phi_{\rm ext}}{\partial r^2},
\end{equation}
i.e. \cref{eq:dispersion}, identical to equations (9) and (10) of \citet{Jog:2013yza}. As in that work, the relation is one-dimensional, governing modes with $\mathbf k \parallel \hat{\mathbf r}$; the position-dependent term $\nabla\rho_1 \cdot \nabla\Phi_{\rm ext}$, which slowly distorts a mode rather than amplifying it, is neglected in this local approximation (see section~4 of \citealt{Jog:2013yza} for its range of validity).

\section{Derivation of the tidal field of the companion galaxy}\label{app:tidal field}

We derive the radial component of the tidal field generated by the companion galaxy starting from its gravitational potential. The gravitational potential of a point mass $M_c$ at distance $R_c$ from the cloud is given by \cref{eq:comp-pot}, where the expression of $R_{c}$ is given in \cref{eq:Rc}.

\subsection{Expansion of the potential}\label{App: taylor expansion}

Since the gas cloud is embedded within the host galaxy, its distance from the host centre is much smaller than the separation between the galaxies, i.e. $r \ll D(t)$. Under this condition, we expand the potential about $\mathbf r = 0$.

We first write
\begin{equation}
R_c = |\mathbf r - \mathbf D| = \sqrt{D^2 + r^2 - 2\,\mathbf r \cdot \mathbf D}.
\end{equation}
Factoring out $D$, this becomes
\begin{equation}
R_c = D \sqrt{1 + \frac{r^2}{D^2} - 2\frac{\mathbf r \cdot \mathbf D}{D^2}}.
\end{equation}

Defining
\begin{equation}
\epsilon = \frac{r^2}{D^2} - 2\frac{\mathbf r \cdot \mathbf D}{D^2},
\end{equation}
thus, $R_{c} = D\sqrt{1+\epsilon}$. So, we can write
\begin{equation}
\frac{1}{R_c} = \frac{1}{D}(1+\epsilon)^{-1/2}.
\end{equation}

Using the binomial expansion
\begin{equation}
(1+\epsilon)^{-1/2} \approx 1 - \frac{1}{2}\epsilon + \frac{3}{8}\epsilon^2 - \cdots,
\end{equation}
and retaining terms up to second order in $r/D$, we obtain
\begin{equation}
\frac{1}{R_c} \approx
\frac{1}{D}
+ \frac{\mathbf r\cdot\mathbf D}{D^3}
- \frac{r^2}{2D^3}
+ \frac{3(\mathbf r\cdot\mathbf D)^2}{2D^5}.
\end{equation}
Multiplying by $-GM_c$, the gravitational potential becomes \cref{eq:comp-pot-big}. As mentioned in the main text, we will be proceeding with the third term.

\subsection{Radial component of the tidal field}\label{app: full_tidal}

The tidal tensor is defined by \cref{eq:tidaltensor}. To obtain the radial component, we project along the radial direction,
\begin{equation}
\tau_{rr} = \hat{r}_i \hat{r}_j \tau_{ij}.
\end{equation}

Writing the tidal potential (the quadrupole term of \cref{eq:comp-pot-big}) as
\begin{equation}
\Phi_{\rm tidal}(\mathbf r) = -\frac{GM_c}{2D^5}\left[3(\mathbf r\cdot\mathbf D)^2 - r^2 D^2\right],
\end{equation}
its gradient is
\begin{equation}
\frac{\partial \Phi_{\rm tidal}}{\partial x_i} = -\frac{GM_c}{D^5}\left[3(\mathbf r\cdot\mathbf D)\, D_i - D^2 x_i\right],
\end{equation}
so that
\begin{equation}
\frac{\partial^2 \Phi_{\rm tidal}}{\partial x_i\, \partial x_j} = -\frac{GM_c}{D^5}\left[3 D_i D_j - D^2 \delta_{ij}\right].
\end{equation}
The tidal tensor, \cref{eq:tidaltensor}, is therefore
\begin{equation}
\tau_{ij} = \frac{GM_c}{D^3}\left(3\hat D_i \hat D_j - \delta_{ij}\right),\label{eq:full-tensor}
\end{equation}
which is symmetric and trace-free, with eigenvalues $+2GM_c/D^3$ along $\hat{\mathbf D}$ and $-GM_c/D^3$ (doubly degenerate) perpendicular to it. Projecting onto the cloud's radial direction,
\begin{equation}
\begin{split}
    \tau_{rr} = \hat r_i \hat r_j\, \tau_{ij} &= \frac{GM_c}{D^3}\left[3(\hat{\mathbf r}\cdot\hat{\mathbf D})^2 - 1\right] \\&= \frac{GM_c}{D^3}\left(3\cos^2\theta - 1\right),
\end{split}
\end{equation}
which is \cref{eq:tidal-radial-comp}. The angular dependence arises from the quadrupolar nature of the tidal field generated by the companion galaxy.

From \cref{eq:parabolic-separation}, we can write, 
\begin{equation}
    \frac{1}{D^3} = \bigg(\frac{1+\cos f}{2r_p}\bigg)^3 
\end{equation}

Using the identity,
\begin{equation}
    1+\cos f = 2\cos^2\bigg(\frac{f}{2}\bigg) 
\end{equation}
we obtain the radial tidal field component of the companion galaxy as,
\begin{equation}
    \tau^{\rm comp}_{rr}(t) =
\frac{GM_c}{r_p^3}
\cos^6\left(\frac{f(t)}{2}\right)
\left(3\cos^2\theta - 1\right).
\end{equation}
which matches \cref{eq:comp-rr} exactly in the main text.

\section{Comparison of the free-fall timescale and the timescale of the companion passage}\label{app:timescales}

To assess whether the tidal perturbation associated with the galaxy encounter can dynamically influence cloud collapse, we compare the characteristic free-fall timescale of representative interstellar clouds with an approximate companion passage timescale.

The free-fall time is given by
\begin{equation}
t_{\rm ff}=\sqrt{\frac{3\pi}{32G\rho_0}},\label{eq:free-fall time}
\end{equation}
where $\rho_0$ is the cloud mass density \citep{GalacticDynamics-book}. 

We define the companion passage time as the duration for which the companion remains within $D \le 2r_p$ of the host, where $r_p$ is the pericentre distance. Substituting $D = 2r_p$ into \cref{eq:parabolic-separation} gives $\cos f = 0$, i.e. $f = \pm 90^\circ$, and the passage time then follows from \cref{eq:barker} as
\begin{equation}
t_{\rm passage} = 2\,|t - t_p|,
\end{equation}
the factor of two accounting for the symmetric approach and recession of the companion about pericentre.

Using Barker's equation with the adopted parameters ($r_p=12~{\rm kpc}$, $M_{\rm host}=10^{12}~M_\odot$, and $M_c=10^{11}~M_\odot$), we obtain
\begin{equation}
|t-t_p|=35.2~{\rm Myr},
\end{equation}
which gives an estimated companion passage time
\begin{equation}
t_{\rm passage}=2|t-t_p|\approx70.4~{\rm Myr}.
\end{equation}
Comparing this value with the free-fall times listed in \cref{tab:cloud-properties} shows that, for the WNM ($t_{\rm ff}\approx58.2~{\rm Myr}$), the cloud collapse timescale is comparable to but smaller than the companion passage time. Thus, the tidal perturbation persists long enough for the cloud to dynamically respond during the close encounter. For denser clouds, the free-fall time is significantly shorter than the passage time, indicating that gravitational collapse proceeds much more rapidly than the duration of the interaction.

This comparison also yields a direct estimate of the delay in the onset of star formation: a diffuse (WNM) region destabilised near pericentre completes its collapse only $\sim t_{\rm ff} \approx 58$~Myr after the passage, comparable to the encounter duration itself, so tidally assisted star formation in diffuse gas lags the pericentric passage by roughly a free-fall time, whereas in dense gas ($t_{\rm ff} \lesssim 1$--$5$~Myr) any tidal modulation of the threshold is expressed essentially instantaneously on the orbital timescale.

\begin{figure}
    \centering
    \includegraphics[width=1\linewidth]{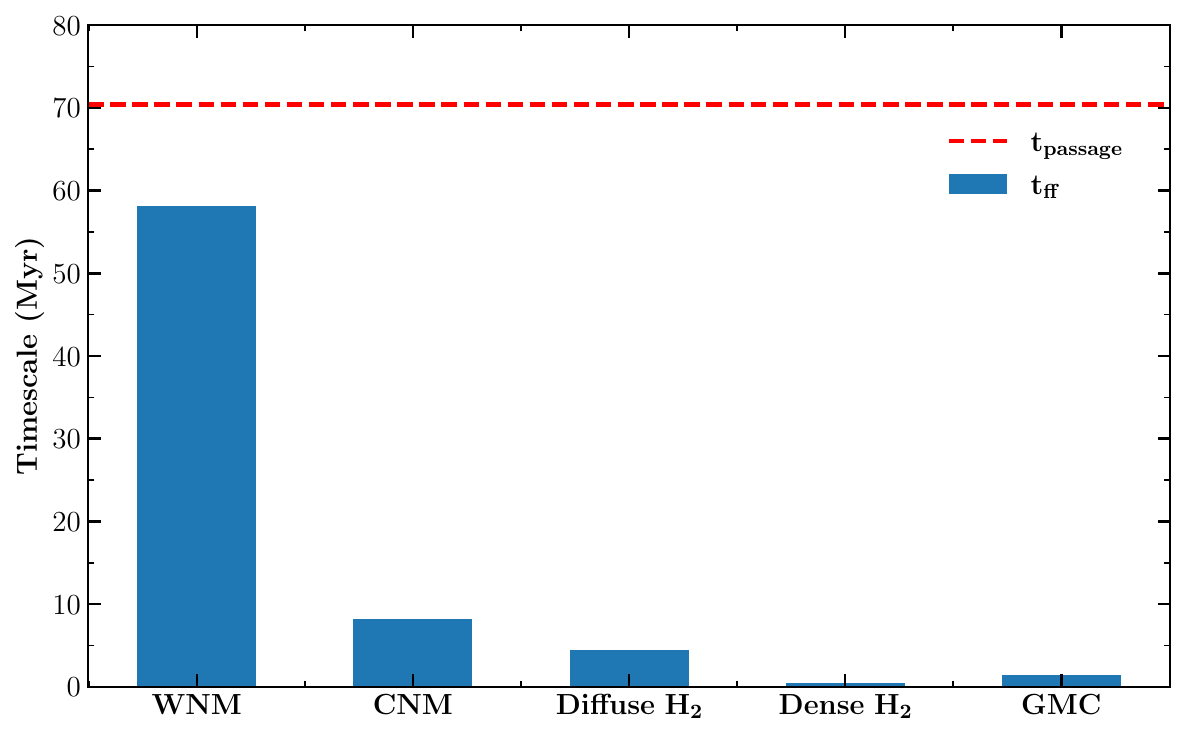}
    \caption{Comparison between the free-fall timescale ($t_{\rm ff}$) of representative interstellar clouds and the estimated companion passage time ($t_{\rm passage}\approx70.4$ Myr). The dashed horizontal line denotes the estimated duration of the companion's close passage ($D\le2r_p$).}
    \label{fig:tff-tpassage}
\end{figure}

\begin{figure}
    \centering
    \includegraphics[width=1\linewidth]{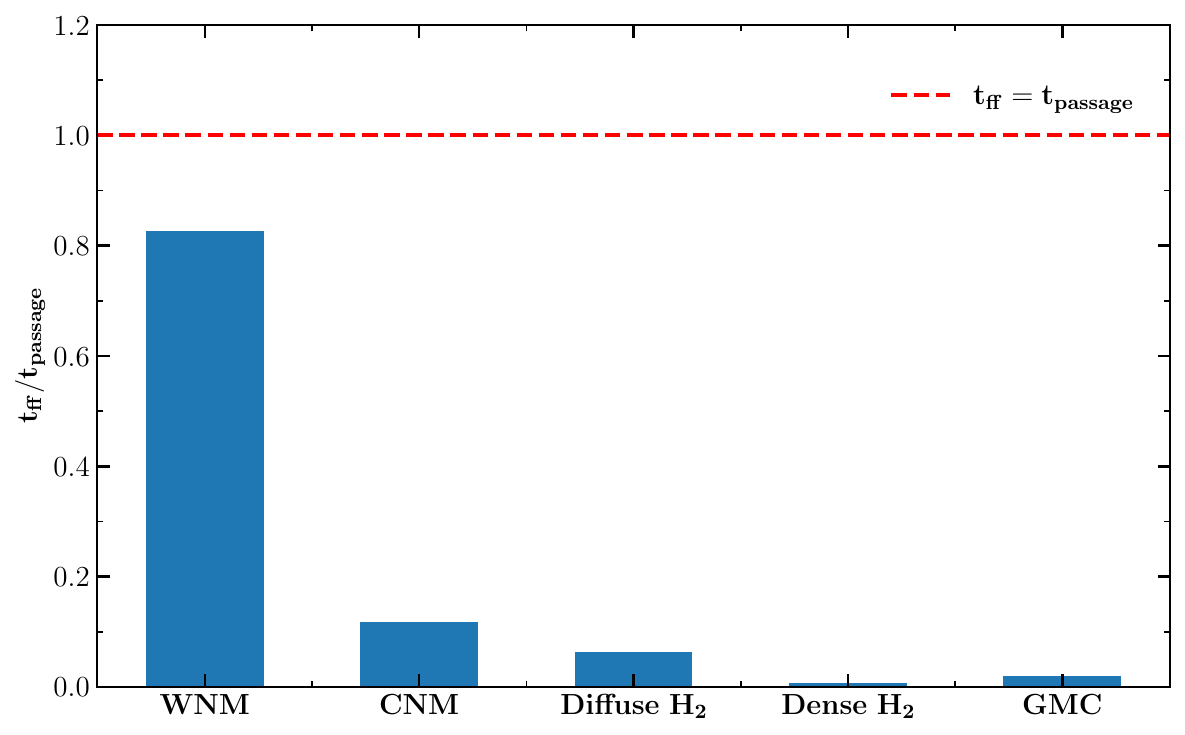}
    \caption{Ratio of the cloud free-fall time to the estimated companion passage time for the representative interstellar cloud models considered in this work.}
    \label{fig:tff-ratio-tpassage}
\end{figure}

\begin{figure}
    \centering
    \includegraphics[width=1\linewidth]{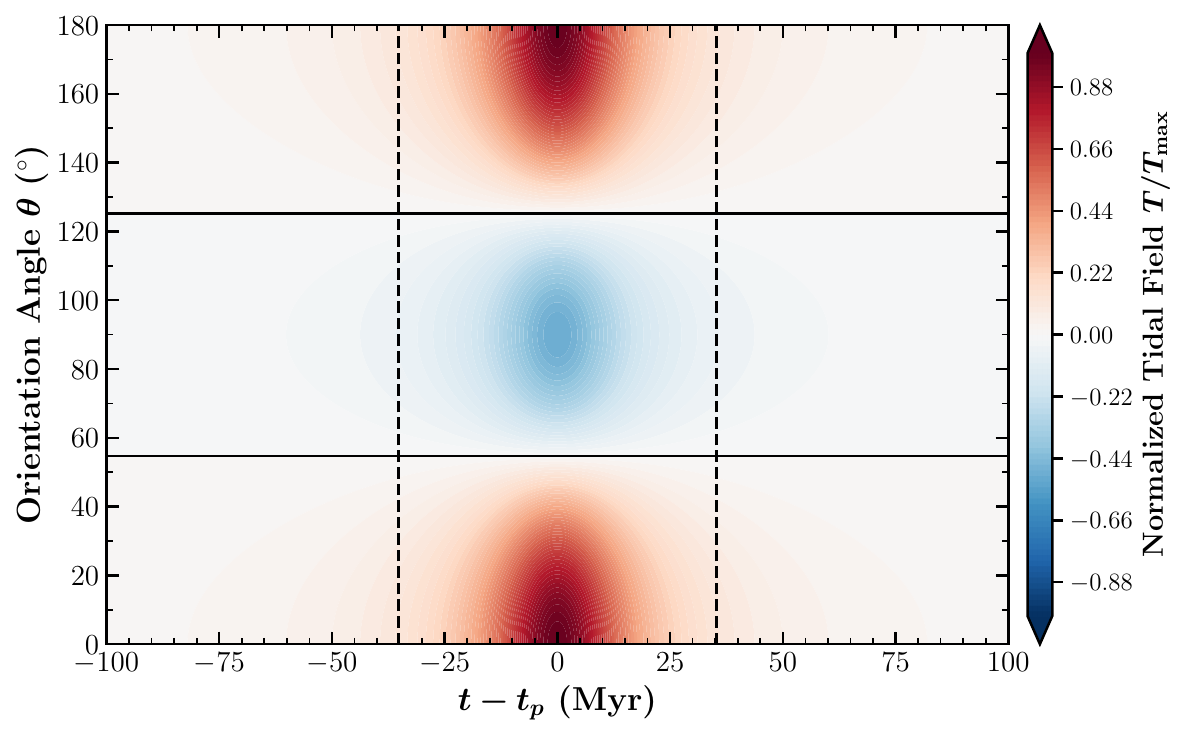}
    \caption{Normalized radial tidal field produced by the companion galaxy as a function of time relative to pericentre ($t-t_p$) and orientation angle ($\theta$). The tidal field is computed using \cref{eq:comp-rr} and normalized by its maximum absolute value. Positive values (red) correspond to disruptive tidal fields, while negative values (blue) indicate compressive tidal fields. The vertical dashed lines mark the interval corresponding to $D\le2r_p$, adopted as the approximate companion passage time, while the horizontal solid lines at $\theta \approx 54.7^\circ$ and $125.3^\circ$ denote the transition between disruptive and compressive tidal regimes ($T=0$).}
    \label{fig:tcomp-time-theta-plot}
\end{figure}

\cref{fig:tff-tpassage}--\cref{fig:tcomp-time-theta-plot} provide complementary diagnostics of the characteristic timescales associated with cloud collapse and the temporal evolution of the companion-induced tidal field during the galaxy encounter. These figures are intended to illustrate the physical interpretation of the analytical results presented in the main text.

As shown in \cref{fig:tff-tpassage} and \cref{fig:tff-ratio-tpassage}, $t_{\rm ff} < t_{\rm passage}$ for every phase: the companion remains near pericentre long enough for the clouds to respond dynamically to the tidal perturbation. The WNM is the marginal case, with $t_{\rm ff}/t_{\rm passage} \approx 0.8$, while the denser phases collapse well within the encounter.

To visualize the evolution of the companion-induced tidal field throughout the encounter, \cref{fig:tcomp-time-theta-plot} presents the normalized tidal field as a function of time relative to pericentre and orientation angle.

\cref{fig:tcomp-time-theta-plot} illustrates the temporal evolution of the companion-induced tidal field for different orientation angles. As expected from the $D^{-3}$ dependence, the tidal field reaches its maximum magnitude near pericentre ($t=t_p$), where the companion is at its minimum separation from the host galaxy, and decreases rapidly as the companion moves away. The figure also highlights the angular dependence of the tidal field, with compressive ($T<0$) and disruptive ($T>0$) regimes separated by the critical angles $\theta \approx 54.7^\circ$ and $125.3^\circ$.

\end{document}